\documentclass[aip,graphicx]{revtex4-1}
\usepackage{graphicx}
\graphicspath{{Figures/}}
\usepackage{array}
\usepackage{amsmath}
\usepackage{bm}

\draft 

\begin{document}


\title{Two-stream instability with a growth rate insensitive to collisions in a dissipative plasma jet
} 



\author{Yi Zhou}
\email[Author to whom correspondence should be addressed: ]{yzhou3@caltech.edu}
\author{Paul M. Bellan}
\affiliation{Department of Applied Physics and Materials Science, California Institute of Technology, Pasadena, California 91125, USA}


\date{\today}

\begin{abstract}
The two-stream instability (Buneman instability) is traditionally derived as a collisionless instability with the presumption that collisions inhibit this instability. We show here via a combination of a collisional two-fluid model and associated experimental observations made in the Caltech plasma jet experiment, that in fact, a low frequency mode of the two-stream instability is indifferent to collisions. Despite the collision frequency greatly exceeding the growth rate of the instability, the instability can still cause an exponential growth of electron velocity and a rapid depletion of particle density. High collisionality nevertheless has an important effect as it enables the development of a double layer when the cross-section of the plasma jet is constricted by a kink-instigated Rayleigh-Taylor instability.
\end{abstract}


\maketitle 

\section{Introduction}
The two-stream instability, also known as the Buneman instability, is a fundamental plasma behavior that can lead to rapid, unstable growth of small perturbations, resulting in effective dissipation of currents in plasmas~\cite{buneman1958instability}. This instability is typically triggered when the electron drift velocity relative to ions is faster than the electron thermal velocity~\cite{buneman1959dissipation}.
The two-stream instability is believed to be related to the formation of a double layer (DL)~\cite{Carlqvist1973,block1978double,quon1976formation,iizuka1979buneman,Ergun2002,charles2007review,mozer2013megavolt} which is a large, localized electric field parallel to the current flow or magnetic field inside a plasma. This localized electric field is called a double layer because the associated charge density given by Poisson's equation consists of two spatially-separated and oppositely-charged layers of particles~\cite{block1978double}.
Studying the two-stream instability and double layers is of value in understanding particle energization in astrophysical plasmas~\cite{alfven1986double}, nuclear fusion~\cite{thonemann1958controlled}, ~\cite, plasmas for space propulsion~\cite{charles2009plasmas}, and laboratory plasma discharges~\cite{takeda1991observations}.

Traditionally, the two-stream instability is derived from two-fluid equations or Vlasov equations by neglecting collision terms\cite{buneman1958instability,buneman1959dissipation,anderson2001tutorial}, and it is commonly presumed that the inclusion of collision terms damps the instability.
In fact, many analytical and numerical studies~\cite{cottrill2008kinetic,singhaus1964beam,hao2009relativistic,self1971growth,sydorenko2016effect} have argued that collisions suppress the instability. 
However, these studies typically do not account for the momentum change of ions as a result of collisions, resulting in violation of momentum conservation.
In this paper, we present a collisional two-fluid model which conserves total momentum.
The momentum conservation enables this two-fluid model to describe a very low frequency two-stream instability which, contrary to conventional presumptions, maintains its characteristic behavior even if the plasma is extremely collisional.
This low frequency two-stream instability will be referred to as an evacuation instability because it has similarities to an evacuation mechanism proposed long ago by Alfv{\'e}n and Carlqvist~\cite{alfven1967currents} and then elaborated by Carlqvist~\cite{carlqvist1972formation} in the context of density depletion and double layer formation with the important exceptions that here (i) collisionality is taken into account and (ii) the driver of the evacuation instability is a naturally occurring periodic constriction of the plasma cross-section.

\section{Observations in the Caltech plasma jet experiment}
\begin{figure}
    \centering
    \includegraphics[scale=1.2]{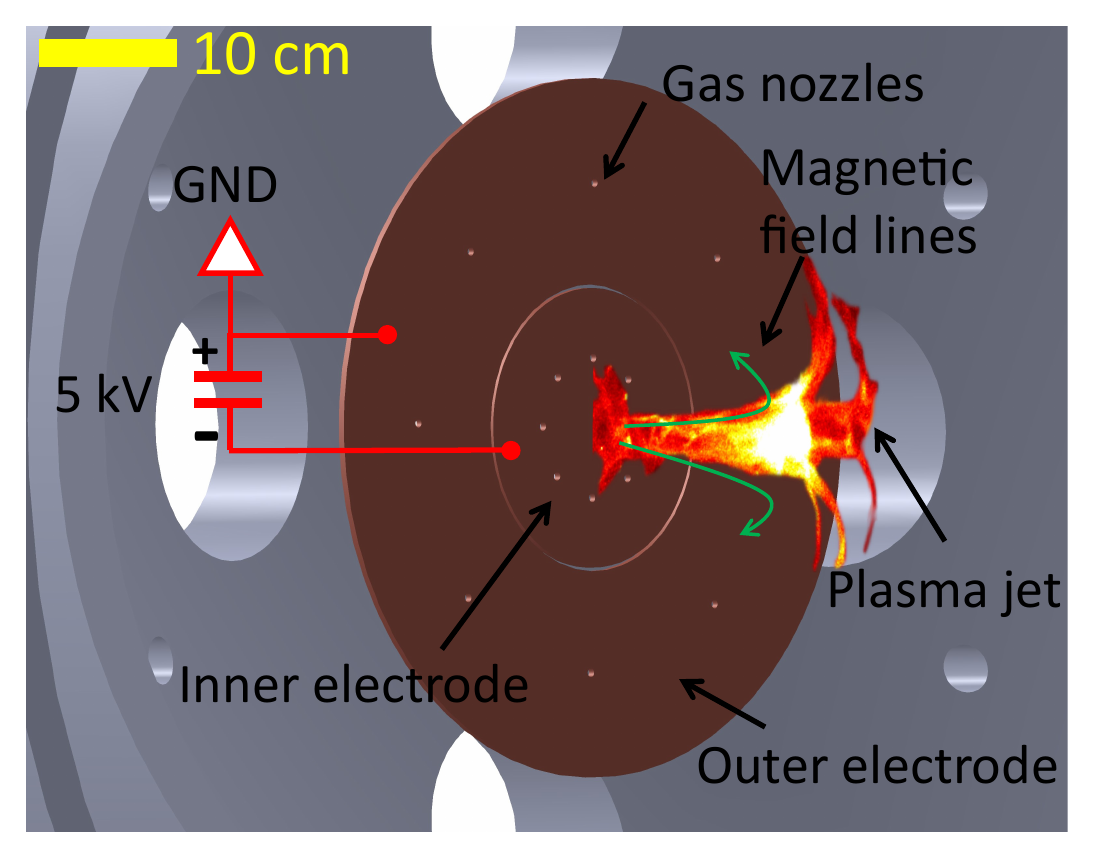}
    \caption{Setup of the Caltech plasma jet experiment with photo of one stage of plasma jet superimposed. A pair of coplanar, concentric electrodes in a large vacuum chamber launch the collimated argon plasma jet shown in the photo. The green magnetic field lines are created by a coil located behind the inner electrode. Neutral gas puffs are injected through 8 pairs of gas nozzles evenly spaced on the electrodes, and then a capacitor bank charged to 5 kV is switched across the inner and outer electrodes to break down the neutral gas. Initially, 8 plasma arches that follow the magnetic field lines form. These 8 arches then merge into the collimated plasma jet shown in the photo.}
    \label{fig:setup}
\end{figure}

The motivation of this study is to explain previous extreme ultraviolet (EUV) and visible light observations made in the Caltech plasma jet experiment~\cite{chai2016extreme}.
The experiment setup, illustrated in Figure~\ref{fig:setup},
is described in detail elsewhere~\cite{hsu2003experimental,you2005dynamic, kumar2009nonequilibrium, moser2012magnetic,chai2016extreme,seo2017spatially,bellan2018experiments}, and here we describe some key features. 
This experiment launches a magnetohydrodynamically-driven (MHD-driven), cold (2 eV), dense ($10^{22}$ $\text{m}^{-3}$) argon plasma jet from a pair of coplanar, concentric electrodes placed inside a large vacuum chamber.
Just before the plasma jet is launched, a coil coaxial with and located behind electrodes generates a dipole-like magnetic field (0.03--0.06 T).
Neutral argon gas puffs are then transiently injected from 8 pairs of gas nozzles evenly spaced  on the electrodes.
After the nozzles have injected a gas cloud localized near the electrodes, a capacitor bank charged to 5 kV is switched across the electrodes to ionize the gas cloud. The capacitor bank then drives an electric current that flows between the electrodes along eight plasma arches that follow the dipolar magnetic field lines linking the electrodes~\cite{you2005dynamic}.
These eight arches then merge into a collimated plasma jet whose initial radius is a few cm. 
This plasma jet then lengthens over time into regions well beyond the extent of the initial gas cloud.
Fig.~\ref{fig:setup} illustrates this collimated plasma jet after it has lengthened to around 20 cm.
The current is further sustained by a set of capacitors and inductors connected to form a pulse forming network.
Except for the initial 10 $\mu$s, the current is maintained to be approximately 70 kA during the 40--50 $\mu$s duration of the plasma jet, and the voltage across the electrodes varies between 2 to 3 kV. Since the plasma density in the jet volume is many orders of magnitude larger than the density of argon gas that was present immediately before the plasma jet was created, the jet cannot be considered as a discharge in a pre-existing argon atmosphere.

\begin{figure}
\includegraphics[scale=0.5]{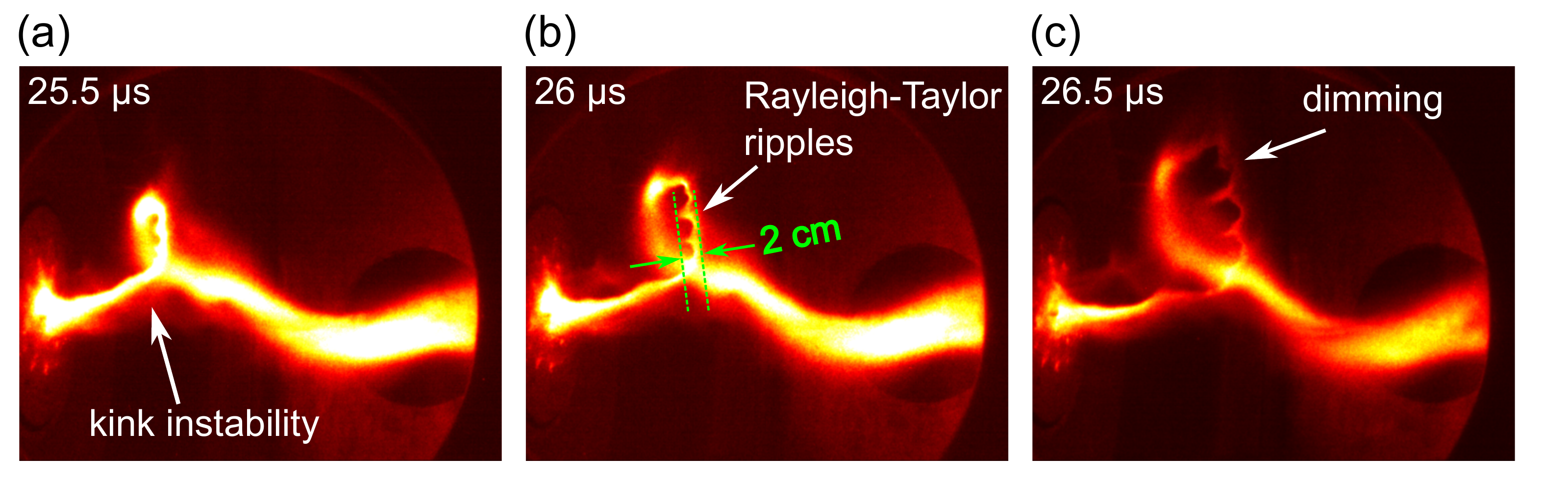}
\caption{\label{fig:plasmaJet} A sequence of false color images of the plasma jet in the Caltech experiment. (a) The plasma jet becomes helical due to the Kruskal-Shafranov kink instability. (b) -- (c) The current channel cross-section becomes constricted at the location of the Rayleigh-Taylor ripples and then dims.}
\end{figure}

Figure~\ref{fig:plasmaJet}
shows a typical sequence of transient events that occur after the jet has lengthened from several centimeters to tens of centimeters. At approximately 20 $\mu$s the jet develops a helical instability (kink), and Fig.~\ref{fig:plasmaJet}(a) shows this kink after it has grown in about 5 $\mu$s to a finite size; the kink identification and its onset being consistent with Kruskal-Shafranov theory   was presented previously~\cite{hsu2003experimental}.
Since kinking  is a rapid exponential growth of a corkscrew shape, kinking produces a strong
lateral acceleration of each segment of the   plasma jet away from the   initial axis of the jet. This lateral acceleration provides a large effective gravity ($10^{10} -10^{11}$ m  s$^{-2}$) that causes a secondary instability 
 called the Rayleigh-Taylor (RT)
instability; this rapid acceleration and the resulting RT instability ripples were reported previously~\cite{moser2012magnetic}.   Fig.~\ref{fig:plasmaJet}(a) and (c) show that in 1 $\mu$s these ripples significantly constrict the cross-section of the plasma jet. Chai, Zhai and Bellan~\cite{chai2016extreme} observed that the region constricted by the RT ripples became bright in EUV radiation (20--60 eV, see Fig. 3 of Chai, Zhai and Bellan~\cite{chai2016extreme}) but, as can be seen in Fig.3 of Chai, Zhai and Bellan and  in  Fig.~\ref{fig:plasmaJet}(c) here, the constricted region also became dim in visible light. The visible light dimming suggests a reduction of the plasma density $n$ has occurred since visible light emission is proportional to  $n^2$, 
the EUV radiation suggests localized plasma heating.
It has been unclear why the dimming in visible light and the brightening in EUV should occur simultaneously. 
\section{Model Description}
\subsection{Overview of the model}\label{sec:overview}
\begin{figure}
\includegraphics[scale=0.5]{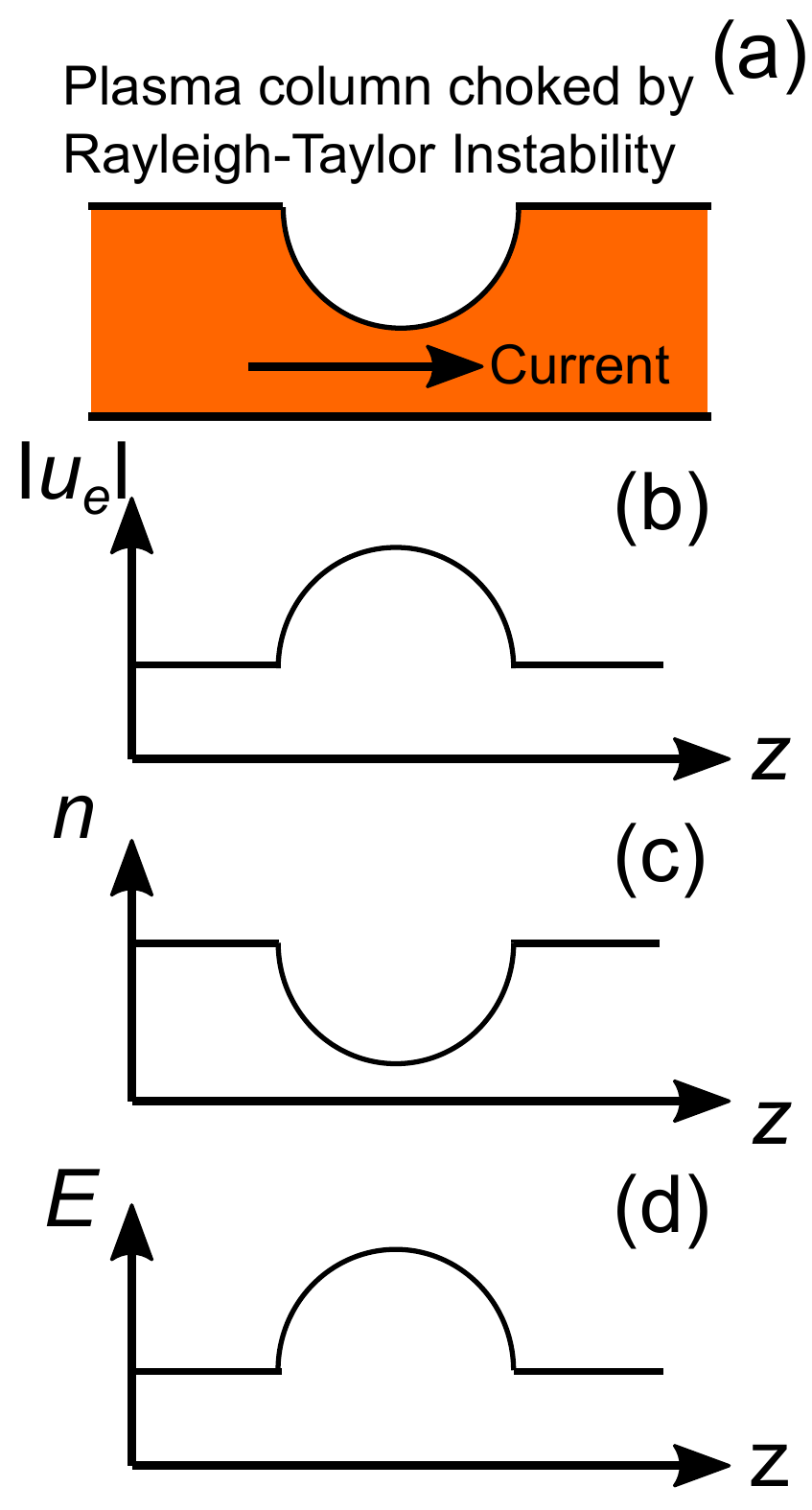}
\caption{\label{fig:dE} The cartoon in (a) illustrates the choking of the plasma jet current channel cross-section $A$ by RT ripples as shown in Fig.~\ref{fig:plasmaJet}(b). Plots (b)--(d) illustrate the electron drift velocity $|u_e|$, density $n$, and electric field $E$ at the constricted location respectively.}
\end{figure}

We present here a collisional two-fluid model  consistent with the observed behavior; this model describes an evacuation instability that has a rapid growth  rate $\gamma$ despite $\gamma \ll \nu_{ei}$, where $\nu_{ei}$ is the electron-ion collision frequency. The model critically depends on the plasma jet having a large electric current flowing along the jet axis   ($z$ direction).
Since magnetic forces are perpendicular to currents, there are no magnetic forces in the $z$ direction, so a 1-D electrostatic model  describes the dynamics along the $z$ direction.
The jet current $I$ is constant because the power supply driving the jet   operates in a constant current mode. This means that  the current density $J=I/A$ must increase when the RT ripples constrict the jet cross-sectional area $A$. MHD instabilities such as the kink and RT are incompressible~\cite{NEWCOMB1960232} and so imply  the  density $n$ is constant when the plasma jet is perturbed by the kink and RT instabilities.
Because $J= nq_eu_e $, where $q_e$ is the electron charge and $u_e<0$ is the electron drift velocity relative to ions, $u_e$ should become more negative when $J$ increases.
Hence, the absolute value $|u_e|$ will peak at the constricted location, as shown in Fig.~\ref{fig:dE} (a) and (b).
 Fig. \ref{fig:plasmaJet}(a) and (c) show that the radius of the plasma jet current channel is reduced from initially greater than 1 cm to eventually less than 0.3 cm. As a result, $A=\pi r^2$ is constricted approximately by a factor of 10 to 20 (from  $>3$ cm$^2$ to $<0.2$ cm$^2$). With $I=70$ kA, $n=10^{22}$ m$^{-3}$, $|u_e|=I/Anq_e$ can increase from below $1\times10^5$ m/s to above $2\times 10^6$ m/s.  The electron thermal velocity of the 2 eV plasma jet is approximately $v_{Te}=\sqrt{\kappa T_e/m_e}\approx6\times10^5$ m/s. Therefore, the initially slower electron drift velocity will greatly exceed $v_{Te}$ due to the constriction of $A$.
It will be shown that this suprathermal electron drift velocity triggers the evacuation instability that can create a density cavity at the constricted location, as shown in Fig.~\ref{fig:dE}(c).
This density cavity would then cause the dimming of visible light.

Due to the combination of low temperature ($T_e\approx T_i\approx 2\ \text{eV}$)  and high density ($n_e\approx n_i\approx 10^{22}\ \text{m}^{-3}$), the characteristic electron-ion collision frequency for the plasma jet \cite{Richardson2019} is
\begin{equation}
\nu_{ei}=2.9\times10^{-12}\frac{n_i\ln{\Lambda}}{T_e^{3/2}}\approx 10^{11}\ \text{s}^{-1},\label{eq:collision}
\end{equation}
where $\ln\Lambda$ is the Coulomb logarithm whose value is assumed to be around 10, and $T_e$ is in units of electron volts. This fast collision frequency corresponds to a $10^{-10}$ s -- $10^{-11}$ s characteristic collision time scale.
Since this collision time is  five to six orders of magnitude smaller than the $10^{-6}$ s -- $10^{-5} $ s jet dynamic time scale, the  jet dynamics is highly collisional. 
Thus, a model explaining how the RT ripples lead to simultaneous visible light dimming (density evacuation) and EUV brightening must take into account that there are $10^{5}$  - $10^{6} $ electron-ion collisions in  the  1 $\mu$s observed time scale of the visible light dimming and EUV brightening. Clearly whatever is happening cannot be described by a collisionless instability.
One may argue that the subset of suprathermal electrons at the constricted region could be collisionless because the electron-ion collision frequency for these fast electrons should be significantly smaller~\cite{Dreicer1959run}. 
A quick estimate of this smaller collision frequency shows that this argument is not correct:
we can roughly estimate the collision frequency by replacing  the $T_e$ in Eq.~(\ref{eq:collision}) with the kinetic energy of electrons traveling at $u_e\approx 2\times10^6$ m/s, i.e., $m_eu_e^2/2\approx 11$ eV. This kinetic energy reduces the collision frequency to be around $8\times 10^{9}\ \text{s}^{-1}$, which corresponds to a $1\times10^{-10}$ s characteristic collision time.

Because the plasma jet is extremely collisional, we can consider the plasma jet as a resistor whose resistance is inversely proportional to $A$.
When $A$ is constricted by RT ripples, the resistance at the constricted location should increase.
As the current carried by the plasma jet flows through this location with larger resistance, a DL, as shown in Fig.~\ref{fig:dE}(d), would develop to heat the plasma locally. This heating can potentially explain the observed EUV radiation.
A collisional two-fluid model that contains the presumptions discussed above will be derived in the next section.

\subsection{Derivation of the collisional two-fluid model} \label{assumptions}
We can derive our collisional two-fluid model by neglecting various small terms from the familiar 1-dimensional, unmagnetized, two-fluid equations for a fully ionized, collisional plasma, namely 
\begin{align}
    \frac{\partial n_{e}}{\partial t}+\frac{\partial}{\partial z}(n_{e}u_{e})&=0,\label{eq:e_con}\\
    \frac{\partial n_{i}}{\partial t}+\frac{\partial}{\partial z}(n_{i}u_{i})&=0,\label{eq:i_con}\\
     m_e\left(\frac{\partial u_e}{\partial t}+u_e\frac{\partial u_e}{\partial z}\right)&= q_e E-\frac{1}{n_e}\frac{\partial P_e}{\partial z}-\frac{R_{ei}}{n_e},\label{eq:motion}\\
          m_i\left(\frac{\partial u_i}{\partial t}+u_i\frac{\partial u_i}{\partial z}\right)&= q_i E-\frac{1}{n_i}\frac{\partial P_i}{\partial z}-\frac{R_{ie}}{n_i},\label{eq:i_motion}\\
    \varepsilon_0\frac{\partial E}{\partial z}&=n_iq_i+n_eq_e\label{eq:gauss}.
\end{align}
Variables in the above equations are defined in the usual way.
The electron energy equation is assumed to be the isothermal equation of state, i.e., $P_e = n_e\kappa T_e$ with $T_e$ being a constant, and the ion energy equation is assumed to be the adiabatic equation of state $P_i\propto n_i^\Gamma$ where $\Gamma$ is an adiabatic constant of order unity. Justification of these energy equations can be found below. The collision terms satisfy $R_{ei}+R_{ie}=0$ because the collisions between electrons and ions must conserve the overall momentum.  The derivation of Eqs.~(\ref{eq:e_con}--\ref{eq:gauss}) can be found in several plasma physics textbooks~\cite{bellan2008fundamentals,nicholson1983introduction,fitzpatrick2015plasma}.

For simplicity, we consider the two-fluid equations in a reference frame that moves with the plasma jet center of mass velocity. In this frame, ions are nearly stationary because the mean ion velocity approximately equals the plasma jet center of mass velocity.
As a result, $u_e$ in this frame not only describes the mean electron drift velocity but also the electron drift velocity relative to ions.
This use of $u_e$ is consistent with the definition of $u_e$ in section~\ref{sec:overview}.

We will analyze equations near an unstable equilibrium with an initial density $n_0$ and an initial suprathermal drift velocity $u_{e0}$ that satisfies $1\ll  u_{e0}/v_{Te}\ll\sqrt{m_i/m_e}$.
This fast $u_{e0}$ is achieved when the plasma jet cross-section is significantly constricted.
This equilibrium is perturbed by an evacuation instability that has a space-time dependence proportional to $\exp(ik_z z+\gamma t)$ where the wavenumber $k_z= 2\pi/\lambda$ is defined by the wavelength $\lambda$ of the RT ripples ($\sim$ 1 cm) and the growth rate $\gamma$ is assumed to be faster than that of the RT instability ($\gamma_{RT}$ is experimentally measured to be on the order of $1\times10^6$ s$^{-1}$) but slower than $k_zv_{Te}$. 

\begin{figure}
    \centering    \includegraphics[scale=0.7]{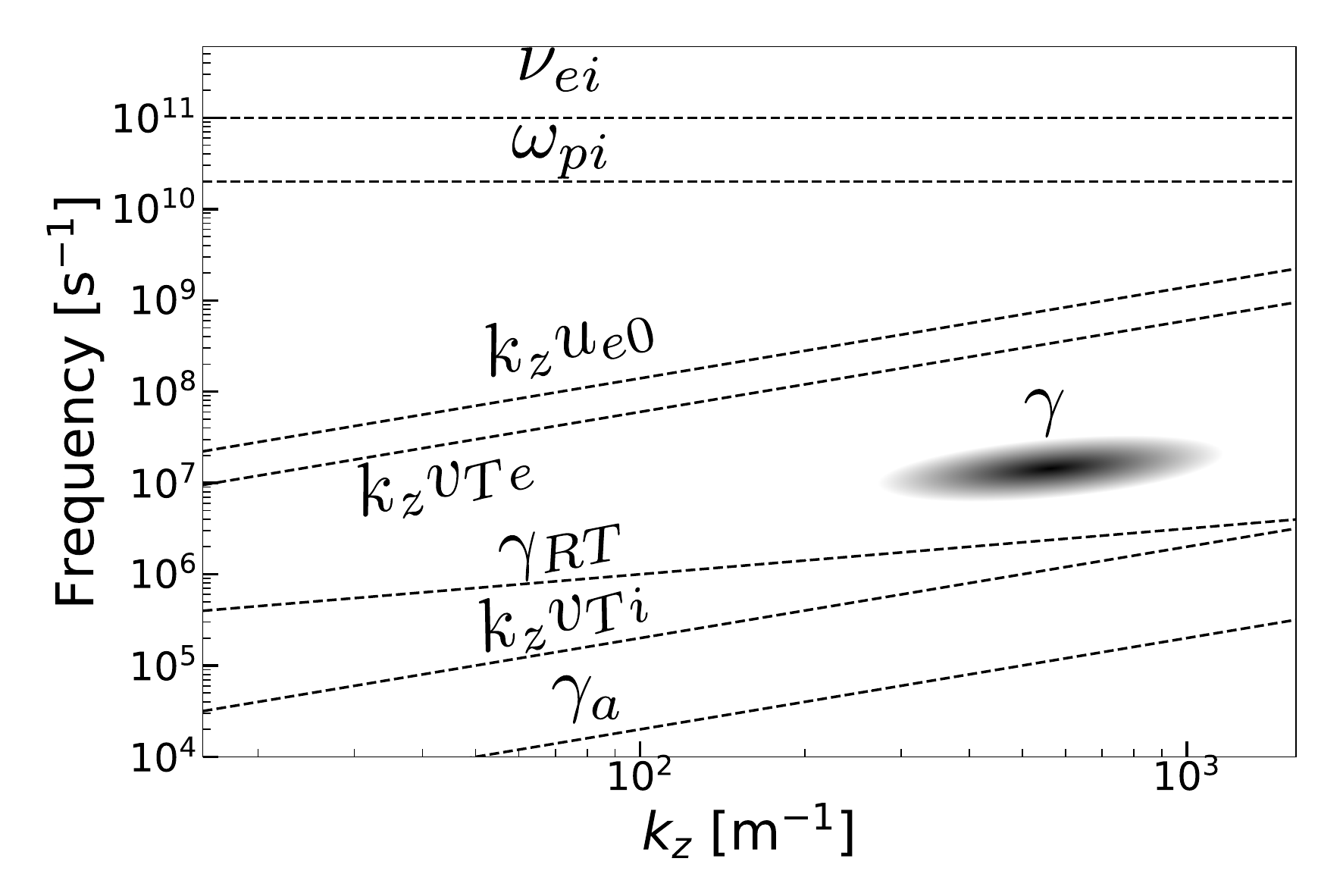}
    \caption{A plot of various characteristic frequencies in the Caltech plasma jet as a function of $k_z$. The dashed lines represent characteristic frequencies estimated with experimental measurements. The growth rate $\gamma_{RT}$ of the RT instability is approximated by the formula $\gamma_{RT}\approx\sqrt{g_\text{eff}k_z}$, where the effective gravity $g_{\text{eff}}$ is estimated from image data to be $\sim 10^{10}$ m/s$^{2}$. The ion acoustic instability growth rate $\gamma_a$ is approximated to be an order of magnitude smaller than $k_z c_s$ where $c_s$ is the ion acoustic velocity $\sqrt{\kappa T_e/m_i}$ (see section~\ref{sec:ion_a}). The growth rate $\gamma$ of the evacuation instability, indicated by the shaded ellipse, is assumed to be at a intermediate location that is either much bigger or smaller than a characteristic frequency. This separation in frequencies enables several simplifications of the original two-fluid equations.}
    \label{fig:frequency_bd}
\end{figure}

Fig.~\ref{fig:frequency_bd}
shows the assumed $\gamma$ is either much smaller or much bigger than several characteristic frequencies relevant to the Caltech plasma jet experiment. 
After an expression for $\gamma$ is derived, it can be directly verified that $\gamma$ indeed falls in the range indicated by this plot, i.e.,
\begin{equation}
\gamma_a\ll k_zv_{Ti}\ll\gamma_{RT}\ll\gamma\ll k_zv_{Te}\ll k_z u_{e0}\ll\omega_{pi}\ll \nu_{ei}\label{eq:ineq},
\end{equation} 
where $\gamma_a$ is the growth rate of the ion-acoustic instability (see section~\ref{sec:ion_a}), and $\omega_{pi}$ is the ion-plasma frequency.
The physically relevant inequality~(\ref{eq:ineq}) establishes that a number of terms in Eqs.~(\ref{eq:e_con})--(\ref{eq:gauss}) are extremely small and so may be neglected; the detailed arguments for dropping these terms are as follows: 
\begin{enumerate}
\item The plasma is quasi-neutral, so $n_e=n_i=n$ when ions are singly charged. From now on, $n$ will be used to denote either the electron or ion density. This simplification is equivalent to assuming the left-hand side (LHS) of Eq.~(\ref{eq:gauss}) is negligible compared to either one of the two terms on the right-hand side (RHS), i.e.,
\begin{equation}
    \left|\varepsilon_{0}\frac{\partial E}{\partial z}\right|\ll\left|n_{i}q_{i}\right|,\left|n_{e}q_{i}\right|.
\end{equation}
so the two terms on the RHS approximately balance each other. 

Justifying this quasi-neutral simplification requires the wavelength of the perturbation to be much longer than the electron Debye length $\lambda_{De}=\sqrt{\varepsilon\kappa T_e/nq_e^2}$. This requirement can be written as
\begin{equation}
    k_z^2\lambda_{De}^2\ll 1,
\end{equation}
which is satisfied when $k_zv_{Ti}\ll\gamma\ll\omega_{pi}$ because
\begin{equation}
    k_z^2\lambda_{De}^2=\frac{k_z^2v_{Te}^2}{\omega^2_{pe}}=\frac{k_z^2v_{Ti}^2}{\omega^2_{pi}}\ll\frac{\gamma^2}{\omega^2_{pi}}\ll 1.
\end{equation}
\item The partial time derivative in Eq.~(\ref{eq:e_con}) is negligible compared to the spatial derivative of $n$, i.e.,
\begin{equation}
    \left|\frac{\partial n}{\partial t}\right|\ll\left|u_e\frac{\partial n}{\partial z}\right|\label{ieq:2}.
\end{equation}
Ignoring the time derivative simplifies Eq.~(\ref{eq:e_con}) to be $\nabla \cdot (nu_e\hat{z})=0$ which when
integrated over the volume of a current channel extending from $z_1$ to $z_2$, as shown in Fig.~\ref{fig:flux_tube},
\begin{figure}
    \includegraphics[scale=0.5]{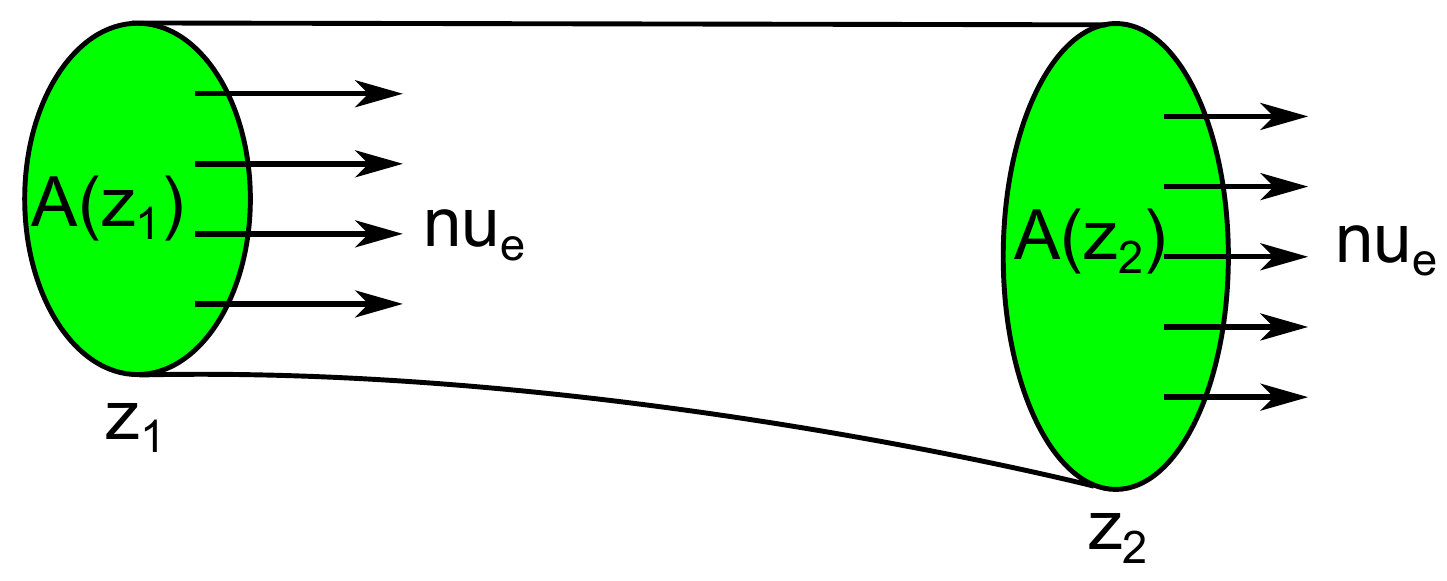}
    \caption{A current channel with varying cross-sectional area extending from $z_1$ to $z_2$. The product $n u_e A$ is independent of $z$.}
    \label{fig:flux_tube}
\end{figure}
gives $n(z_2)u_e(z_2)A(z_2)-n(z_1)u_e(z_1)A(z_1)=0$. Since $z_1$ and $z_2$ are arbitrary, the equation above is equivalent to $n(z)u_e(z)A(z)=\text{constant}$ for any $z$ (specifically, $nu_eA$ is constant at the constriction location). This corresponds to stating that the electric current flowing along the channel is independent of $z$.

This simplification is justified when $\gamma$ is small compared to $k_zu_{e0}$. This is because the linearized form of inequality~(\ref{ieq:2}) can be reduced to  $\gamma\ll k_zu_{e0}$ due to the $\exp{(ik_z z+\gamma t)}$ dependence of perturbations.

\item Similar to simplification 2, the partial time derivative in Eq.~(\ref{eq:motion}) can be ignored compared to the convective term, i.e.,
\begin{equation}
    \left|\frac{\partial u_{e}}{\partial t}\right|\ll\left|u_{e}\frac{\partial u_{e}}{\partial z}\right|.\label{ieq:3}
\end{equation}

\item Electrons are isothermal, so the electron pressure is $P_e=n\kappa T_e$ where $T_e$ is a constant.

The validity of this assumption relies on $\gamma$ being much smaller than $k_z{v_{Te}}$. This is because electrons can be considered as isothermal for a perturbation with a characteristic velocity $\gamma /k_z$ that is slow compared to the electron thermal velocity $v_{Te}$.

\item The ion pressure term can be ignored compared to the partial time derivative in Eq.~(\ref{eq:i_motion}), i.e.,
\begin{equation}
    \left|\frac{1}{n}\frac{\partial P_{i}}{\partial z}\right|\ll \left|m_{i}\frac{\partial u_{i}}{\partial t}\right|\label{ieq:5}.
\end{equation}
This simplification is equivalent to assuming $k_zv_{Ti}\ll \gamma $ because
the the linearized form of inequality~(\ref{ieq:5}) can be rewritten as
$\Gamma k_{z}^{2}v_{Ti}^{2}\ll\gamma^{2}$ and the adiabatic constant $\Gamma$ is of order unity.
To see how this can be done, we first 
multiply both sides of inequality~(\ref{ieq:5}) by $n_0/m_i$ and express $P_{i1}$ in terms of $n_{1}$ using the linearized adiabatic equation of state $P_{i1}= \Gamma\kappa T_{i}n_{1}$:
\begin{equation}
     \left|\Gamma v_{Ti}\frac{\partial n_{1}}{\partial z}\right|\ll \left|n_0\frac{\partial u_{i1}}{\partial t}\right|.
\end{equation}
We can then differentiate both sides with respect to $z$ and eliminate $u_{i1}$ using the linearized form of Eq.~(\ref{eq:i_con}) (see Eq.~[\ref{eq:ui1}]) to get
\begin{equation}
        \left|\Gamma v_{Ti}\frac{\partial^2 n_{1}}{\partial z^2}\right| \ll \left|\frac{\partial^2 n_{1}}{\partial t^2}\right|.
\end{equation}
Assuming $n_1$ is proportional to $\exp{(ik_z z+\gamma t)}$ gives us desired inequality $\Gamma k_{z}^{2}v_{Ti}^{2}\ll\gamma^{2}$.
\end{enumerate}

The 5 simplifications listed above reduce the original system of two-fluid equations to be
\begin{align}
    nq_eu_eA&=I=\text{constant},\label{eq:e_continuity}\\
    \frac{\partial n}{\partial t}+\frac{\partial}{\partial z}(n u_i)&=0,\label{eq:ion_continuity}\\
    m_eu_e\frac{\partial u_e}{\partial z}&= q_e E-\frac{\kappa T_e}{n}\frac{\partial n}{\partial z}-\frac{R_{ei}}{n},\label{eq:e_motion}\\
    m_i\left(\frac{\partial u_i}{\partial t}+u_i\frac{\partial u_i}{\partial z}\right)&= q_i E-\frac{R_{ie}}{n}\label{eq:ion_motion}.
\end{align}
Adding Eqs.~(\ref{eq:e_motion}) and (\ref{eq:ion_motion})  gives 
\begin{align}
m_{e}u_{e}\frac{\partial u_{e}}{\partial z}
+m_{i}\left( \frac{\partial u_{i}}{\partial t}+u_{i}\frac{\partial u_{i}}{
\partial z}\right) +\frac{\kappa
T_{e}}{n}\frac{\partial n}{\partial z}=0.  \label{eq:sum_electron_ion_motion}
\end{align} which has the interesting feature of being a two-fluid equation that, while taking collisions into account, does not explicitly depend on collisions. By two-fluid, it is meant that the equation is beyond the scope of MHD as there is an explicit dependence on electron mass. The collision terms in Eqs.~(\ref{eq:e_motion}) and (\ref{eq:ion_motion}) can be arbitrarily large without affecting Eq.~(\ref{eq:sum_electron_ion_motion}).

If the collision effects are neglected and the cross-sectional area $A$ is assumed to be uniform, the system of Eqs.~(\ref{eq:e_continuity})--(\ref{eq:sum_electron_ion_motion}) reduces to Carlqvist's collisionless evacuation mechanism~\cite{carlqvist1972formation} and to the collisionless equations studied by Galeev et al.~\cite{galeev1981nonlinear} and by Bulanov and Sarosov~\cite{bulanov1986ion}.
Because collision effects and a constricted $A$ are not considered in Refs.~\onlinecite{carlqvist1972formation,galeev1981nonlinear,bulanov1986ion}, these studies failed to describe the unidirectional DL electric field shown in Fig.~\ref{fig:dE}(d) (note that the electric field in  Fig. 2d of Carlqvist~\cite{carlqvist1972formation} is bidirectional). 
It will be shown in section~\ref{field_section} that collision effects and a constricted $A$ are critical for producing the unidirectonal DL electric field.

\subsection{Growth rate of the evacuation instability}
We now consider the cross-section $A$ to vary on the time scale of the RT instability which is assumed to be much slower than the evacuation instability being derived. Thus, $A$ is a slowly varying parameter from the point of view of the evacuation instability, and a decrease of $A$ with a consequent increase of $u_{e}$ is effectively the ``knob'' that triggers the fast evacuation instability. Because the decrease of $A$ is caused by the RT instability, the evacuation instability can be considered as a tertiary instability triggered by the secondary RT instability.

The growth rate $\gamma$ of the evacuation instability can be calculated in the jet frame (ion velocity is nearly zero in this frame) by first linearizing  Eqs.~(\ref{eq:e_continuity}), (\ref{eq:ion_continuity}), and (\ref{eq:sum_electron_ion_motion}) about an initial equilibrium with a uniform density $n_0$ and an electron drift velocity $u_{e0}$ to obtain
\begin{align}   n_0u_{e1}+n_1u_{e0}&=0,\label{eq:le_continuity}\\
    \frac{\partial n_1}{\partial t}+n_0\frac{\partial u_{i1}}{\partial z}&=0,\label{eq:lion_continuity}\\
    m_eu_{e0}\frac{\partial u_{e1}}{\partial z}+m_i\frac{\partial u_{i1}}{\partial t}+\frac{\kappa T_e}{n_0}\frac{\partial n_1}{\partial z}&=0\label{eq:lsum},
\end{align}
where the subscript 1 denotes a small perturbation to the initial equilibrium.
The cross-sectional area $A$ is treated as a constant when linearizing Eq.~(\ref{eq:e_continuity}).
We can express $u_{e1}$ and $u_{i1}$ in terms of $n_1$ using Eqs.~(\ref{eq:le_continuity}) and (\ref{eq:lion_continuity}):
\begin{align}   
u_{e1}&=-\frac{n_1u_{e0}}{n_0}\label{eq:ue1},\\
\frac{\partial u_{i1}}{\partial z}&=-\frac{1}{n_0}\frac{\partial n_1}{\partial t}\label{eq:ui1}.
\end{align}
Multiplying Eq.~(\ref{eq:lsum}) by $-n_0$ and differentiating the resulting equation with respect to $z$ give
\begin{equation}
    -n_0m_eu_{e0}\frac{\partial^2 u_{e1}}{\partial z^2}-n_0m_i\frac{\partial^2 u_{i1}}{\partial t\partial z}-\kappa T_e\frac{\partial^2 n_1}{\partial z^2}=0.
\end{equation}
We can get an equation that only involves $n_1$ by eliminating $u_{i1}$ and $u_{e1}$ using Eqs.~(\ref{eq:ue1}) and~(\ref{eq:ui1}):
\begin{equation}
m_eu_{e0}^2\frac{\partial^2 n_1}{\partial z^2}+m_i\frac{\partial^2 n_1}{\partial t^2}-\kappa T_e\frac{\partial^2 n_1}{\partial z^2}=0.
\end{equation}
We can then derive a dispersion relation by assuming the density perturbation $n_1$ has an $\exp{(ik_z z+\gamma t)}$ dependence:
\begin{equation}
    -k_z^2m_eu_{e0}^2+\gamma^2m_i+k_z^2\kappa T_e=0,
\end{equation}
which can be solved for $\gamma$ to obtain
\begin{equation}
    \gamma=k_z\sqrt{\frac{m_e}{m_i}(u_{e0}^2-v_{Te}^2)}\label{eq:growth}.
\end{equation}

Equation~(\ref{eq:growth}) shows $\gamma$ is real and positive when $u_{e0}$ is greater than $v_{Te}$, so this instability is triggered when reduction of $A$ causes the electron drift velocity  to exceed the electron thermal velocity.  It is important to note that, while collisions have been taken into account,  this instability does not depend on whether or not the plasma is collisional. Thus, the instability should occur in a highly collisional plasma such as the Caltech plasma jet. Using $k_z=700\ \text{m}^{-1}$, $m_e=9.1\times10^{-31}$ kg, $m_i=6.7\times10^{-26}$ kg,  $u_{e0}= 2\times 10 ^6 $ m/s, and $v_{Te}=6\times 10^{5}$ m/s, we can estimate the instability growth rate $\gamma$ for the Caltech argon plasma jet to be $\gamma\approx 5\times10^{6}\ \text{s}^{-1} $, which is indeed consistent with Fig.~\ref{fig:frequency_bd}.

\subsection{Electric field}\label{field_section}
The electric field in the plasma jet can be found from Eq.~(\ref{eq:motion}):
\begin{equation}
     E=\frac{m_e}{q_e}\left(\frac{\partial u_e}{\partial t}+u_e\frac{\partial u_e}{\partial z}\right)+\frac{1}{nq_e}\frac{\partial P_e}{\partial z}+\frac{R_{ei}}{nq_e},\label{eq:E}
\end{equation}
As discussed in section~\ref{sec:overview} and shown in Figure~\ref{fig:frequency_bd}, the electron-ion collision frequency $\nu_{ei}$ is much larger than any other characteristics frequencies, such as $k_{z}u_{e0}$ and  $k_zv_{Te}$.
The collisional term $R_{ei}/nq_e$ is proportional to $\nu_{ei}$:
\begin{equation}
    \frac{R_{ei}}{nq_e}=\frac{\nu_{ei}m_e(u_e-u_i)}{q_e} \approx \frac{\nu_{ei}m_eu_e}{q_e}.
\end{equation}
 Thus, we can assume the collision term in Eq.~(\ref{eq:E}) is the dominant term that balances the electric field, i.e.,
\begin{equation}
    E\approx\frac{\nu_{ei}m_eu_e}{q_e}.\label{eq:E_approx}
\end{equation}
This electric field is in fact a DL that is inversely proportional to the cross-sectional area $A$.
Before the plasma jet is constricted significantly, the electron drift velocity is slower than the electron thermal velocity $v_{Te}$. To first order, we can assume $\nu_{ei}$ is initially proportional to $n T_e^{-3/2}\ln\Lambda $. Since $T_e$ is a constant for isothermal electrons and $\ln{\Lambda}$ does not vary much, $\nu_{ei}$ is approximately proportional to $n$. Thus, the electric field in Eq.~(\ref{eq:E_approx}) is approximately proportional to the electron flux $nu_e$, which is inversely proportional to the cross-sectional area $A$ because $nu_e=I/q_e A$. Therefore, this electric field is enhanced at the region constricted by RT ripples and appears as a DL as shown in Fig.~\ref{fig:dE}(d).

The combination of this DL and high collisionality of the plasma jet can possibly cause Ohmic heating which can potentially explain the 20--60 eV EUV radiation observed in the plasma jet experiment.  Because the Caltech plasma jet is highly collisional, this DL electric field can be quite large.
For example, this DL can be as large as $\sim6\times10^4$ V/m, when the relative drift velocity is $u_e \approx 1\times 10^{5}$ m/s and the electron-ion collision frequency is $\nu_{ei}\approx 1\times10^{11}\ \text{s}^{-1}$.
A complication associated with Ohmic heating is if the local electron temperature is increased to become sufficiently large, then the condition $ u_e \gg v_{Te}$ would cease and the evacuation instability would be quenched; consideration of this higher order issue will be left for future consideration and so will not be addressed here.

The $1/A$ dependence derived above depends on $\nu_{ei}$ being proportional to $n T_e^{-3/2}\ln\Lambda$. However, assuming $\nu_{ei}$ is proportional to $n T_e^{-3/2}\ln\Lambda$ is not valid when the electron drift velocity becomes significantly faster than the electron thermal velocity~\cite{Dreicer1959run}.
As a result, the DL due to collisions will probably not scale as $1/A$ when the plasma jet is constricted significantly. In fact, the strength of the DL may decrease because the collisionality of suprathermal electrons decreases as the relative drift velocity increases~\cite{Dreicer1959run}. 
When the strength of DL has decreased to a point that it no longer dominates the RHS of Eq.~(\ref{eq:E}), we can retain an extra term from Eq.~(\ref{eq:E}):
\begin{equation}
    E=\frac{R_{ei}}{nq_e}-\frac{m_e}{2|q_e|}\frac{\partial u_e^2}{\partial z}.\label{eq:e_field}
\end{equation}
The newly included term on the RHS of Eq.~(\ref{eq:e_field}) is the leading order correction because the linearized form of this term is proportional to $k_z u_{e0}$, which is smaller than $\nu_{ei}$ but larger than other relevant characteristic frequencies according to Fig.~\ref{fig:frequency_bd}.

The new correction term is proportional to the partial $z$ derivative of $u_e^2$.
Across the constricted region, the velocity profile in Fig.~\ref{fig:dE}(b) has a maximum, so the partial $z$ derivative of $u_e^2$ will be positive on the LHS of the maximum and negative on the RHS of the maximum.
Because the constant in front of the partial $z$ derivative is negative, the second term should point away from constricted region.
Since $u_e$ can grow exponentially once the evacuation instability is triggered, this initially insignificant bidirectional electric field will also grow exponentially and possibly become stronger than the unidirectional DL eventually. Unlike a unidirectional DL, this bidirectional electric field cannot accelerate the bulk of the streaming electrons since the associated electric potential does not have a net jump (the integral of the bidirectional electric field across the constricted region equals 0).
However, a small fraction of electrons can possibly be accelerated by this bidirectional electric field in two opposite directions. These fast electrons could possibly explain why localized X-ray sources near the electrode and far from the electrode have been observed simultaneously in the Caltech jet experiment (see Fig. 9 of Zhou, Pree, and Bellan~\cite{zhou2023imaging}).

\section{Numerical Solution}
A numerical solution of the problem is now presented to demonstrate the evacuation instability when the cross-sectional area of Caltech's plasma jet is significantly constricted by the RT instability. The evacuation instability manifests itself as a fast growth of electron drift velocity and a rapid depletion of density in the numerical solution. This numerical solution replicates and extends the linear analysis.
\subsection{Dimensionless equations and Discretization}
For the numerical treatment, we use Caltech-experiment-relevant reference quantities defined in Table~\ref{tab:refq} to normalize the equations of our collisional two-fluid model. 
\begin{table}
\caption{\label{tab:refq}Definitions of reference quantities relevant to the Caltech plasma jet experiment.}
\begin{ruledtabular}
\begin{tabular}{lll}
    Symbol & Reference quantity & Value\\ [0.5ex]
    \hline
    $z_{\text{ref}}$ & length & 1 cm\\
    $n_{\text{ref}}$ &  density & $10^{22}$ m$^{-3}$\\
    $I_{\text{ref}}$ &  current & 100 kA\\
    $q_{\text{ref}}$ &  charge & $e=1.6021\times10^{-19}$ C\\
    $m_{\text{ref}}$ & mass & $m_e=9.1094\times$$10^{-31}$ kg\\
    $A_{\text{ref}}=z_{\text{ref}}^2$ &  cross-sectional area & 1 cm$^2$\\
    $u_{\text{ref}}=I_{\text{ref}}/A_{\text{ref}}n_{\text{ref}}q_{\text{ref}}$ &  velocity & $6.2\times 10^5$ m/s\\
    $t_{\text{ref}}=z_{\text{ref}}/u_{\text{ref}}$ & time & 16 ns\\
    $T_{\text{ref}}=m_\text{ref}u_\text{ref}^2$ &  Temperature & 2.2 eV
\end{tabular}
\end{ruledtabular}
\end{table}
For the reference quantities in Table~\ref{tab:refq}, the reference length $z_{\text{ref}}$, density $n_{\text{ref}}$, and electric current $I_{\text{ref}}$ are experiment-specific independent quantities that need to be prescribed. Other reference quantities can be derived from these three prescribed quantities, the elementary charge $e$, and the electron mass $m_e$.
With these reference quantities, the dimensionless form of Eqs.~(\ref{eq:e_continuity}), (\ref{eq:ion_continuity}), and (\ref{eq:sum_electron_ion_motion}) are
\begin{align}
    \bar{n}\bar{u}_e\bar{A}&=-\bar{I}=\text{constant},\label{eq:ne_continuity}\\
    \frac{\partial\bar{n}}{\partial\bar{t}}+\frac{\partial}{\partial\bar{z}}(\bar{n}\bar{u}_i)&=0,\label{eq:nion_continuity}\\
    \frac{1}{2}\frac{\partial \bar{u}_e^2}{\partial \bar{z}}+\bar{m}_i\left(\frac{\partial \bar{u}_i}{\partial \bar{t}}+\frac{1}{2}\frac{\partial \bar{u}_i^2}{\partial \bar{z}}\right)+\bar{T}_e\frac{\partial}{\partial \bar{z}}\ln\bar{n}&= 0\label{eq:nion_motion},
\end{align}
where the barred (dimensionless) variables are the original (dimensioned) variables divided by their associated reference quantities. 

The 3 dimensionless equations above are discretized with the Forward Time Centered Space (FTCS) method.
The spatial dimension $\bar{z}$  is discretized over the interval $[-\pi,\pi] $ with a mesh width of $h=\pi/128$. The dimensionless time  $\bar{t}$  is discretized with a discrete time step of $k=0.5h$. 
For any variable $f$ defined on the $\bar{z}-\bar{t}$ space-time plane, $f_j^m$ denotes the pointwise value $f(\bar{z}_j,\bar{t}_m)$. All quantities are assumed to be periodic so that $f(-\pi,\bar{t}_m)=f(\pi,\bar{t}_m)$. The discretized equations read
\begin{align}
\bar{u}_{e,j}^{m}=&\frac{-\bar{I}}{\bar{n}_{j}^{m}\bar{A}_j^m},\label{eq:dis-ue}\\
    \bar{n}_{j}^{m+1}=&\bar{n}_{j}^{m}-\frac{k}{2h}\left(\bar{n}_{j+1}^{m}\bar{u}_{i,j+1}^{m}-\bar{n}_{j-1}^{m}\bar{u}_{i,j-1}^{m}\right),\label{eq:dis-n}\\
    \bar{u}_{i,j}^{m+1}=&\bar{u}_{i,j}^{m}-\frac{k}{2h\bar{m}_{i}}\left[\frac{\left(\bar{u}_{e,j+1}^{m}\right)^{2}-\left(\bar{u}_{e,j-1}^{m}\right)^{2}}{2}+\bar{T}_{e}\left(\ln\bar{n}_{j+1}^{m}-\ln\bar{n}_{j-1}^{m}\right)\right]\nonumber \\
    &-\frac{k}{4h}\left[\left(\bar{u}_{i,j+1}^{m}\right)^{2}-\left(\bar{u}_{i,j-1}^{m}\right)^{2}\right].\label{eq:dis-ui}
\end{align}

\subsection{Numerical results}
\begin{figure}
    \centering
    \includegraphics[scale=3]{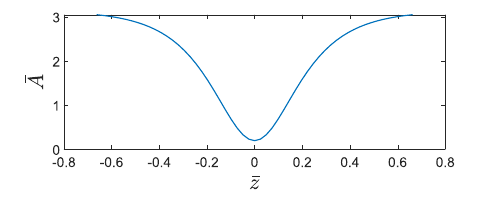}
    \caption{The area function used in the numerical calculation. The area near $\bar{z}=0$ has been reduced significantly to trigger the evacuation instability. }
    \label{fig:area_func}
\end{figure}

Given an area function $\bar{A}$ and parameters $\bar{I}$, $\bar{T_e}$, and $\bar{m_i}$, Eqs.~(\ref{eq:dis-ue})--(\ref{eq:dis-ui}) can be solved recursively for unknowns $\bar{u}_e$, $\bar{n}$, and  $\bar{u}_i$ from an initial ion velocity $\bar{u}^0_{i}$ and an initial density $\bar{n}^0$.
For the numerical solution to be presented in this section, we choose the area function to be $\bar{A}=\pi \left[1-15/16e^{\bar{t}/1000}\text{sech}(2\pi\bar{z})\right]$,
so that the cross-sectional area $\bar{A}$ at $\bar{z}=0$ has been exponentially reduced by RT instability to $\pi/16$ at $\bar{t}=0$.
A plot of this area function at $\bar{t}=0$ is shown in Fig.~\ref{fig:area_func}.
The $e$-folding time is chosen to be 1000, so that the area is essentially constant on the time scale of the evacuation instability.
The dimensionless parameters $\bar{I}$, $\bar{T_e}$, and $\bar{m_i}$ are chosen to be 0.7, 1, and 70000 respectively to match the conditions in the plasma jet experiment.
For the initial condition, we use an initially uniform density profile $\bar{n}^0_j=1$ and an initially stationary ion velocity profile $\bar{u}^0_{i,j}=0$.

\begin{figure}
    \includegraphics[scale=3]{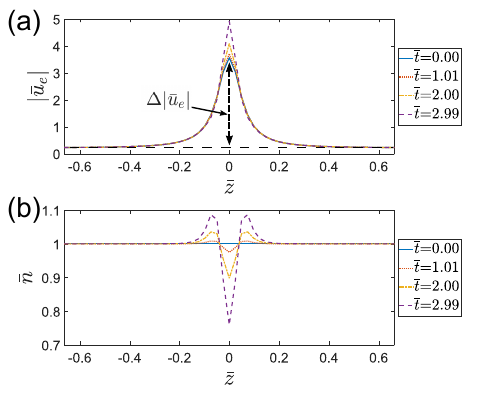}
    \caption{The numerically calculated (a) electron drift velocity and (b) density at different times around $\bar{z}=0$. The electron drift velocity becomes faster and the density dip becomes deeper due to the evacuation instability triggered by the constriction of the plasma jet cross-section. $\Delta |\bar{u}_e|$ in (a) denotes the change in electron drift velocity. Fig.~\ref{fig:growth} shows how $\Delta |\bar{u}_e|$ grows exponentially with time due to the evacuation instability.}
    \label{fig:numerical}
\end{figure}

\begin{figure}
    \centering
    \includegraphics[scale=3]{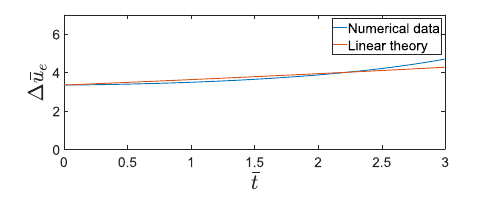}
    \caption{The change in electron drift velocity $\bar{z}=0$ as a function of time is plotted in blue. The red line is the exponential function $3.6e^{0.082\bar{t}}$, where the 0.082 in the exponent is the normalized growth rate calculated from Eq.~\ref{eq:growth}. This normalized growth rate corresponds to a dimensioned growth rate of approximately $5\times 10 ^6$ $\text{s}^{-1}$. }
    \label{fig:growth}
\end{figure}

Fig.~\ref{fig:numerical} shows
the numerically calculated $|\bar{u}_e|$ and $\bar{n}$ around $\bar{z}=0$ at different times.
These numerical results agree with the apparent profiles illustrated in Fig.~\ref{fig:dE}(b) and (c) qualitatively.
In Fig.~\ref{fig:numerical}(a), the absolute value of $\bar{u}_e$ peaks at $\bar{z}=0$ due to the constriction of the plasma jet. $|\bar{u}_e|$ then grows rapidly due to the evacuation instability.
In order to conserve the current $\bar{I}$, a density dip starts developing as shown in Fig.~\ref{fig:numerical}(b).
As time increases, the density dip becomes deeper and deeper.
This density dip is consistent with  the dimming of  visible light observed in the plasma jet experiment.
In Fig.~\ref{fig:growth}, we compare the numerical growth of $|\bar{u}_e|$ to the exponential growth obtained from linear theory. The blue line represents the numerically calculated $\Delta|\bar{u}_e|$ indicated in Fig.~\ref{fig:numerical}, and the red line is an exponential function that grows with a normalized growth rate calculated from Eq.~(\ref{eq:growth}) using initial parameters from the numerical calculation . The numerical solution agrees well with the linear theory until $\bar{t}=2.5$. After $\bar{t}=2.5$, the numerical solution starts to grow faster than the growth obtained from linear theory.

\section{Discussion}
\subsection{Connection to the two-stream instability}
The growth rate $\gamma$ of the evacuation instability is related to a very low frequency approximate solution to the classic dispersion for the two-stream instability~\cite{Carlqvist1973,buneman1958instability}
\begin{equation}
    1-\frac{\omega_{pi}^2}{\omega^2}-\frac{\omega_{pe}^2}{(\omega-k_z u_{e0})^2}=0,\label{eq:two-stream}
\end{equation}
where $\omega$ is the (complex) frequency of the two-stream instability and $u_{e0}\gg v_{Te}$ is implied since there is no mention of $v_{Te}$. 
To obtain the approximate low frequency solution, $\omega_{pi}^2/\omega^2$ and $\omega_{pe}^2/(\omega-k_zu_{e0})^2$ are assumed to be much greater than 1 so that the 1 in Eq.~(\ref{eq:two-stream}) can be ignored; this corresponds to assuming quasi-neutrality. In addition, $\omega$ in the denominator of the electron term in Eq.~(\ref{eq:two-stream}) is neglected compared to $k_z u_{e0}$.
The resulting simplified dispersion 
\begin{equation}
    \frac{\omega_{pi}^2}{\omega^2}+\frac{\omega_{pe}^2}{k_z^2 u_{e0}^2}=0
\end{equation}  
has a positive (unstable) imaginary solution 
$\omega=\text{i}k_zu_{e0}\sqrt{m_e/m_i}$, which is similar to $\gamma$ if $v_{Te}$ is ignored in Eq.~(\ref{eq:growth}). 

However, there is a large difference between the evacuation instability derived here and the classic two-stream instability.
The classic two-stream instability ignores collisions, and many studies have argued that including collisions would suppress the growth rate of the instability~\cite{cottrill2008kinetic,singhaus1964beam,hao2009relativistic,self1971growth,sydorenko2016effect}. 
Contrarily, the derivation of the evacuation instability shows that the existence of the instability, i.e., its threshold and growth rate, is indifferent to collisions since the momentum lost by electrons as a result of collisions equals the momentum gained by ions as demonstrated here in adding Eqs.~(\ref{eq:e_motion}) and (\ref{eq:ion_motion}) to obtain Eq.~(\ref{eq:sum_electron_ion_motion}). 

The evacuation instability's indifference to collisions might have a connection to a new kinetic theory~\cite{tigik2016collisional}, recently studied by Tigik, Ziebell, and Yoon, on collisional damping rates for plasma waves because this kinetic theory has predicted damping rates that are significantly weaker than those according to the traditional theory. However, the situation considered by Tigik, Ziebell, and Yoon has no initial electron drift velocity, i.e., $u_{e0} = 0$ while the situation considered in this paper has a finite $u_{e0}$.
More studies are needed to understand the effect of finite $u_{e0}$ on the collisional damping rates calculated from the newly developed kinetic theory.

\subsection{Difference from the ion-acoustic instability}\label{sec:ion_a}
The ion-acoustic instability is a well-known instability that can be excited in a current-carrying plasma if the electron drift velocity $u_{e0}$ relative to ions is greater than the ion-acoustic velocity $c_s=\sqrt{\kappa T_e/m_i}$. Readers who are familiar with this instability may recall that the standard expression for the ion-acoustic instability growth rate is roughly 
\begin{equation}
    \gamma_a\approx\frac{\pi^{1/2}}{2\sqrt{2}}k_z\left(u_{e0}-c_{s}\right)\sqrt{\frac{m_e}{m_i}},\label{eq:ion_ac}
\end{equation} 
in the limit that $k_z\lambda_{De} \ll 1$ and $T_e\gg T_i$ (see the textbook by Ichimaru~\cite{ichimaru1973basic} for a derivation). If $u_{e0}\gg c_s$, then $\gamma_a\approx k_zu_{e0}\sqrt{m_e/m_i}$, which is seemingly similar in expression to the growth rate $\gamma$ of the evacuation instability (see Eq.~[\ref{eq:growth}]). However,
$\gamma_a$ is assumed to be much smaller than $k_z c_s$ in the calculation that leads to Eq.~(\ref{eq:ion_ac}). In other words, equation~(\ref{eq:ion_ac}) is only valid if $\gamma_a$ is evaluated to be much smaller than $k_z c_s$.
This requirement automatically makes $\gamma_a$ much smaller than $\gamma$ because $u_{e0}\gg v_{Te}$ is assumed in the derivation that leads to $\gamma\approx k_zu_{e0}\sqrt{m_e/m_i}$, so 
\begin{equation}
    \gamma\gg k_z v_{Te}\sqrt{m_e/m_i}=k_z c_s\gg\gamma_a.
\end{equation}
 
\section{conclusion}
In summary, the collisional two-fluid model presented here  extends the well-known two-stream instability to a previously unknown high-collision regime, and this extension predicts an evacuation instability when the current channel cross-section is constricted. 
Applying this model to the Caltech plasma jet experiment demonstrates that electron acceleration, density depletion, and an electric DL associated with localized plasma heating are developed when the plasma jet cross-section is constricted by the RT instability.
Therefore, the collisional two-fluid model is consistent with the visible light dimming and EUV radiation observed in the Caltech plasma jet experiment.
This model is likely applicable to other collisional plasmas, both in nature and in the laboratory, that can develop  constricted cross-sections of current channels.
Examples include solar prominences and the formation of laboratory spheromaks.

\begin{acknowledgments}
Supported by NSF Award 2105492 and by  AFOSR Award FA9550-21-1-0379.
\end{acknowledgments}
\section*{AUTHOR DECLARATIONS}
\subsection*{Conflict of Interest}
The authors have no conflicts to disclose.
\subsection*{Author Contributions}
\textbf{Yi Zhou:}
Formal analysis (equal);
Investigation (equal);
Methodology (equal);
Software (lead);
Validation (equal);
Visualization (lead);
Original draft preparation (lead);
Review and editing (equal).
\textbf{Paul M. Bellan:}
Conceptualization (lead);
Funding acquisition (lead);
Formal analysis (equal);
Investigation (equal);
Methodology (equal);
Project administration (lead);
Supervision (lead);
Validation (equal);
Original draft preparation (supporting);
Review and editing (equal).
\section*{Data availability}
The data that support the findings of this study are available
from the corresponding author upon reasonable request.


%
%

%


\bibliography{References.bib}

\end{document}